\documentclass[11pt]{article}
\usepackage{graphicx}
 \usepackage{mathptmx}    
\usepackage{latexsym}
\usepackage[round]{natbib}
\usepackage{fullpage}
\usepackage{amsmath}
\usepackage{microtype}

\DeclareMathOperator{\SF}{SF}

\DeclareMathOperator{\RF}{RF}
\DeclareMathOperator{\STC}{STC}
\DeclareMathOperator{\LGN}{LGN}
\DeclareMathOperator*{\argmax}{argmax}

\title{\Large \textbf{Development of spatial coarse-to-fine processing in the visual pathway} }
\author{Jasmine A. Nirody \thanks{Biophysics Graduate Group, University of California, Berkeley; Email:  jnirody@berkeley.edu}}
\date{\small July 16, 2013}

\begin{document}

\maketitle

\begin{abstract}
The sequential analysis of information in a coarse-to-fine manner is a fundamental mode of processing in the visual pathway. Spatial frequency (SF) tuning, arguably the most fundamental feature of spatial vision, provides particular intuition within the coarse-to-fine framework: low spatial frequencies convey global information about an image (e.g., general orientation), while high spatial frequencies carry more detailed information (e.g., edges). In this paper, we study the development of cortical spatial frequency tuning. As feedforward input from the lateral geniculate nucleus (LGN) has been shown to have significant influence on cortical coarse-to-fine processing, we present a firing-rate based thalamocortical model which includes both feedforward and feedback components. We analyze the relationship between various model parameters (including cortical feedback strength) and responses. We confirm the importance of the antagonistic relationship between the center and surround responses in thalamic relay cell receptive fields (RFs), and further characterize how specific structural LGN RF parameters affect cortical coarse-to-fine processing. Our results also indicate that the effect of cortical feedback on spatial frequency tuning is age-dependent: in particular, cortical feedback more strongly affects coarse-to-fine processing in kittens than in adults. We use our results to propose an experimentally testable hypothesis for the function of the extensive feedback in the corticothalamic circuit.
\end{abstract}

\section{Introduction} \label{intro}

The mode of information processing in neural sensory systems has been the subject of many experimental and computational studies. It is intuitive that visual information is processed sequentially in a coarse-to-fine manner---when looking quickly at a scene, we process the general features before focusing on individual objects. Indeed, such dynamics have been documented for several tuning parameters in vision, e.g., orientation selectivity (\citealp{ringach1997dynamics,shapley2003dynamics}; but see also \citealp{gillespie2001dynamics, müller2001information,mazer2002spatial}), spatial frequency selectivity \citep{bredfeldt2002dynamics,mazer2002spatial,frazor2004visual,nishimoto2005accuracy} and binocular disparity tuning~\citep{ringach2003look,menz2003stereoscopic}. 

In this paper, we focus on spatial frequency tuning. Spatial frequency (SF) provides special insight into visual perception within the framework of the coarse-to-fine system. Low spatial frequencies convey global information about an image (such as general orientation or proportion), whereas finer detail (such as edges) is provided by higher spatial frequency information \citep{bar2004visual}. In the following, we use the terms \emph{dynamic spatial frequency tuning} and \emph{spatial coarse-to-fine processing} interchangeably. 

The development of spatial coarse-to-fine processing is still an open problem. In mature animals, spatial coarse-to-fine processing was initially thought to be primarily a cortical function, but has recently been observed in the retina and lateral geniculate nucleus (LGN) \citep{enroth1983spatio,allen2006dynamic}. In fact, the spatial coarse-to-fine process in the primary visual cortex (V1) has been shown to be largely due to feedforward contribution from the thalamus in adults \citep{allen2006dynamic}. In particular, cortical SF tuning has been shown to arise from the center-surround organization in LGN spatiotemporal receptive fields \citep{allen2006dynamic}. However, as the structure of these receptive fields (RFs) evolves significantly throughout maturation, more detailed study is necessary to shed light on the development of this process.

We present a model to investigate the development of dynamic SF tuning in the thalamocortical circuit. Using published data from electrophysiological experiments in cats \citep{cai1997spatiotemporal}, we map the RFs of LGN cells from kittens at 4 and 8 weeks postnatal and mature cats. These RFs are used to construct a feedforward-feedback thalamocortical model. We study the response of this model to static sinusoidal grating input in order to investigate how certain features of thalamic spatiotemporal RFs affect feedforward contribution to SF tuning in the cortex.

A key property of the model is its simplicity, which allows us to thoroughly analyze the relationship between its parameters and responses. In addition, because the model includes both feedforward and feedback components, we evaluate how corticothalamic feedback affects the cortical coarse-to-fine dynamic throughout the developmental process. 

A striking feature of our results is that the effect of cortical feedback on coarse-to-fine processing is age-dependent. Specifically, cortical feedback has a stronger effect on spatial frequency tuning early in development. These results point to a role for the large amount of recurrent connections from the visual cortex to the thalamus, the function of which has yet to be agreed upon.

In Section \ref{model}, we provide a thorough description of the structure of the model. We then describe our numerical implementation and reason our choice of model parameters in Section \ref{analysis}. We present our results in Section \ref{results}, focusing in particular on the effects of a few key parameters: relative surround intensity and cortical feedback strength. In Section~\ref{discussion}, we summarize and relate our results to previous findings, and use them to put forward a hypothesis for the function of the extensive thalamocortical feedback in Section~\ref{open}.

\section{Model overview} \label{model}

\begin{figure}
\begin{center}
\includegraphics[width=0.55\textwidth]{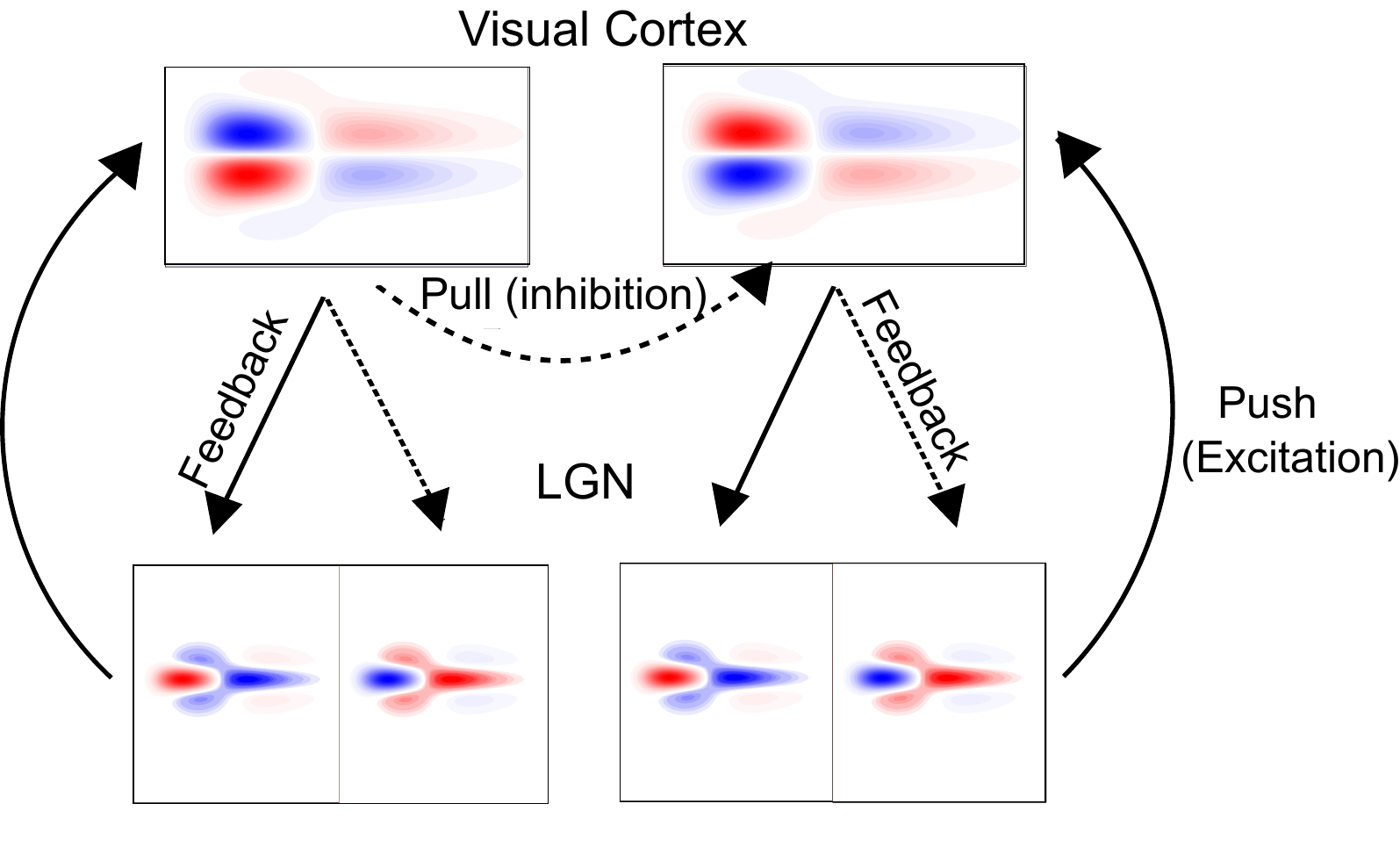}
\caption{Schematic of our feedforward-feedback model. In this figure, cells are represented by their receptive fields (RFs) (see Section~\ref{ss:stim}). Solid and dashed lines indicate excitatory and inhibitory connections, respectively. Thalamic cells (bottom) connected to a cortical cell (top) have overlapping RFs; the offset between LGN RFs has been exaggerated for clarity in this figure. In all spatiotemporal RFs presented in this paper, red and blue contours enclose bright and dark excitatory regions, respectively.}
\label{fig:schematic}
\end{center}
\end{figure}

To study the cortical coarse-to-fine dynamic during development, we construct a phenomenological thalamocortical model. In our model, we measure responses from one excitatory cortical (V1) simple cell. Feedforward connections from the thalamus are modeled explicitly, while corticothalamic feedback is modeled implicitly through modifications to the LGN receptive field. For clarity of presentation, we organize this section as follows. 

We first give an overview of the integration of feedforward input from the LGN by cortical cells (Section \ref{ss:feedforward}). We then detail how this input is generated, first by describing the structure of the LGN population in Section \ref{ss:lgnpop} and then by explaining how the RF of an LGN cell is used to compute a neural response (Section \ref{ss:stim}) and describing in detail the structure of an individual LGN RF (Section \ref{ss:lgn}).  Finally, we explain how this RF structure is modified to simulate cortical feedback in Section \ref{ss:feedback}. 

\subsection{Feedforward input}\label{ss:feedforward}

We construct two cortical simple cells whose RFs are 180$^\circ$ out of phase. This corresponds to having one excitatory and one inhibitory cell. We are primarily concerned with the output of the excitatory cell. This excitatory cell receives input from both the LGN cell population (described in Section~\ref{ss:lgnpop}), as well as from the antiphase cortical cell. The connection from the inhibitory cell functions to imitate inhibitory thalamocortical input. See Figure \ref{fig:schematic} for a basic schematic of the model structure.

\begin{figure*}
\begin{center}
\includegraphics[width=.7\textwidth]{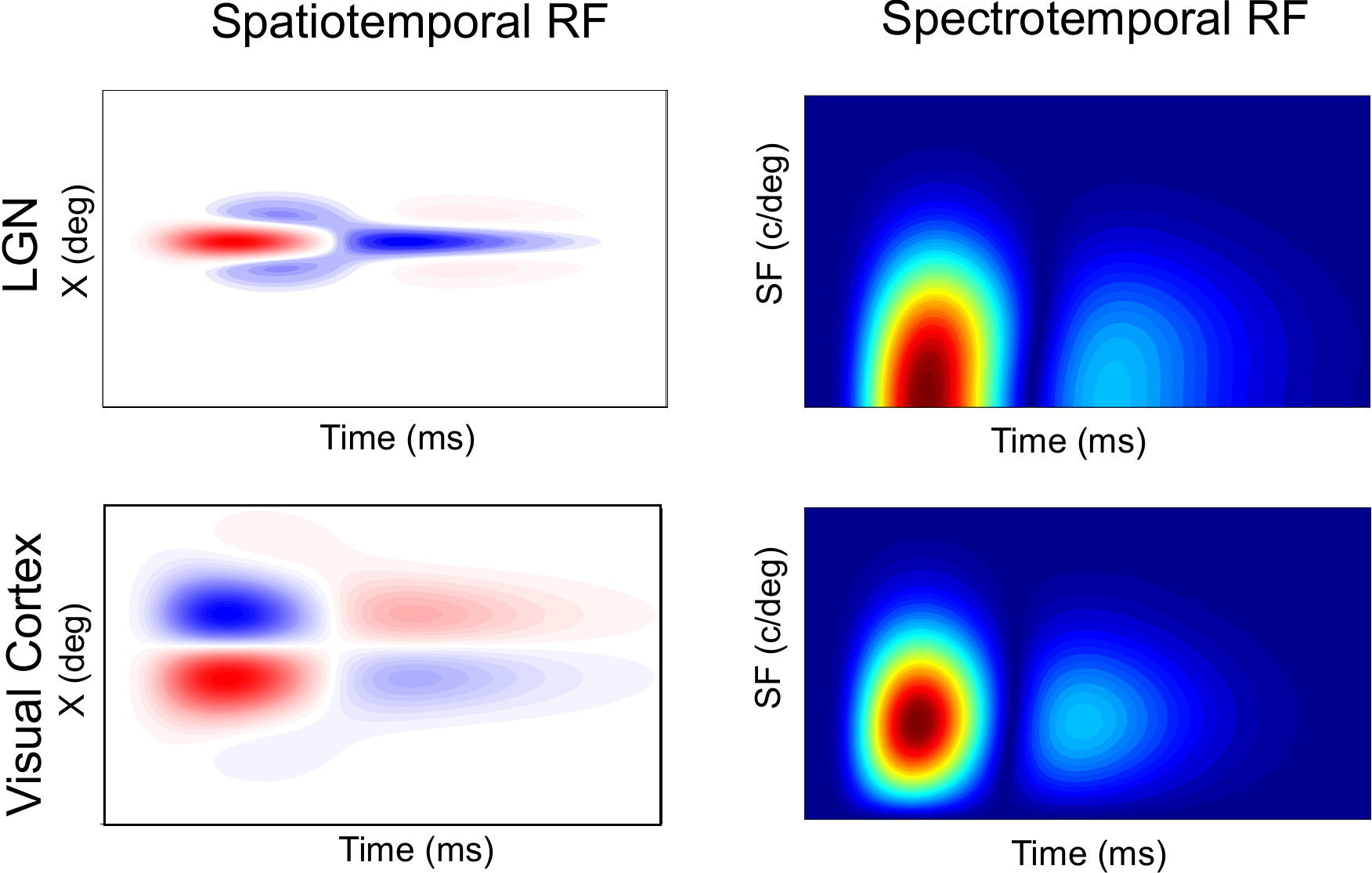}
\caption{Representative spatiotemporal (left) and Fourier-transformed spectrotemporal (right) receptive fields from LGN (top) and visual cortex (bottom). Rightward tilted contours in the spectrotemporal RF indicate the presence of coarse-to-fine processing.}
\label{fig:rfs}
\end{center}
\end{figure*}

Traditionally, cortical simple cells in V1 are modeled using Gabor filters \citep{deangelis1993a}. For simplicity, we model simple cell RFs with two explicit subregions separated by 1$^\circ$, in accordance with data from previous studies \citep{allen2006dynamic}. Flanking subregions are also sometimes present due to the surround response in the LGN. See the bottom left panel of Figure~\ref{fig:rfs}.

Each cortical subregion receives input from $n$ ON- and OFF-center LGN neurons, meaning each cortical cell receives input from $2n$ thalamic cells ($n$ for each of the two subregions). We choose $n=20$, which is within the rough range given by \citet{alonso2001rules} for the number of LGN cells that converge onto a cortical simple cell (see Table~\ref{tab:param} for a full list of parameter values). The response of an LGN cell is given as a function of stimulus frequency $f$, phase $\phi$, and time $t$ (measured in ms); this corresponds to a conversion to the frequency domain using a Fourier transform (see Sections~\ref{ss:stim} and \ref{ss:lgn} for more details). We define inhibitory input as simply a weighted, time-delayed output from a population of LGN cells. Explicitly, the total input to the excitatory cell is given as:
\begin{equation}\label{eq:input}
I(f,\phi,t) = \sum_{j=1}^{2n} \LGN_{e,j}(f,\phi,t) - W \sum_{j=1}^{2n} \LGN_{i,j}(f,\phi,t-\tau).
\end{equation}
Here, $\LGN_e(f,\phi,t)$ and $\LGN_i(f,\phi,t)$ are the outputs of the thalamic neurons connected to the excitatory and inhibitory cortical cells, respectively, $W$ is the weight of the inhibitory connection, and $\tau$ is a delay to account for the transmission of the impulse (see Table~\ref{tab:param}). Details on the computation of $\LGN_e(f,\phi,t)$ and $\LGN_i(f,\phi,t)$ are given in Section \ref{ss:stim}; specifically, an explicit formulation is given in Equation \ref{eq:ftr}.

Given input as in Equation \ref{eq:input}, the output of the excitatory cell is the input integrated with a time constant of 10 ms and rectified \citep{allen2006dynamic}:
\begin{equation}\label{eq:cortresp}
R(f,\phi,t) = \frac{1}{10 \times 2n}\left[ \int_{t-10}^{t} I(f,\phi,t') dt' \right]^{+}.
\end{equation}

\subsection{LGN population structure}\label{ss:lgnpop}

To facilitate easier interpretation of model output, we fix parameters corresponding to the LGN cell population structure. LGN cell receptive fields (Equation \ref{eq:lgnrf}) are positioned at the center of the cortical subregion to which they project. The size of the RF center $\sigma_c$ is equal to the width of the subregions, while the size of the surround is given as $\sigma_{\text{s}} = 1.5 \times \sigma_{\text{c}} + 0.4$~\citep{cai1997spatiotemporal}. 

Since cortical cells receive input from a nonhomogenous LGN cell population, we wish to demonstrate robustness of the model response to variability in parameters controlling cell position and RF center and surround size. To this end, we discuss the randomization of these population parameters in Section \ref{ss:random}.

In the next section, we describe how the RF of an LGN cell is used to generate the output response given an input stimulus.  

\subsection{Stimulus to response}\label{ss:stim}

A spatiotemporal receptive field $\RF(x,t)$ provides the relationship between an input stimulus $S(f,\phi,x)$ of frequency $f$ and phase $\phi$ and the cell response $r(f,\phi, t)$:


\begin{equation*}
r(f,\phi,t) = \int \RF(x,t) S(f,\phi,x) dx.
\end{equation*}
To calculate the firing rate of an LGN cell, we add the spontaneous geniculate firing rate (10 spikes/ms) to this equation and rectify \citep{allen2006dynamic}:
\begin{equation}\label{eq:ftr}
\LGN(f,\phi,t) = \left[r(f,\phi,t) + 10 \right]^{+}.
\end{equation}
This equation describes both excitatory cell ($\LGN_e$) and inhibitory cell ($\LGN_i$) responses. Computations using such predicted spectrotemporal receptive fields have been shown to accurately reproduce directly measured SF tuning dynamics \citep{allen2006dynamic}.

We choose stimulus functions $S(f,\phi,x)$ to be static sinusoidal gratings:
\begin{equation}\label{eq:stim}
S(f,\phi,x) = \cos(2\pi f x-\phi).
\end{equation}
For values of SF $f$ and phase $\phi$ used, see Table~\ref{tab:param}.

In Section \ref{ss:quant}, we explain our choice of metrics to quantify cortical coarse-to-fine processing from V1 cell output. Recall that the integration of LGN output to generate cortical dynamics was described in Section \ref{ss:feedforward}.

The particular structure of the RF of an individual LGN cell is described in detail in the following section.

\subsection{LGN receptive fields}\label{ss:lgn}

LGN cells have receptive fields with a center-surround antagonistic structure \citep{cai1997spatiotemporal}. LGN RFs are typically modeled so as to consist of two components corresponding to the center and surround responses. The peripheral area (``surround'') responds oppositely to light than the central region. We consider two types of LGN cells: ON- and OFF-center cells. ON-center cells are stimulated when light is shined on the center, and inhibited when light is shined on the surround. OFF-center cells display the opposite pattern. See the top left panel in Figure \ref{fig:rfs} for an example of the spatiotemporal RF of an LGN ON-center cell.

We model LGN spatiotemporal RFs based on characterizations from \citet{cai1997spatiotemporal}.  LGN RFs are described by the expression: 
\begin{equation}\label{eq:lgnrf}
\RF(x,t) = F_c(x)G_c(t) - F_s(x)G_s(t),
\end{equation}
where $F(x)$ and $G(t)$ are spatial and temporal profiles, respectively, and subscripts $c$ and $s$ correspond to center and surround responses. Temporal filters are given as difference of gamma functions:
\begin{equation}\label{eq:temporal}
G_c(t) = K_1\frac{c_{1}(t-t_1)^{n_1} e^{-c_{1}(t-t_1)}}{n_1^{n_1} e^{-n_1}} - K_2\frac{c_2(t-t_2)^{n_2} e^{-c_{2}(t-t_2)}}{n_2^{n_2} e^{-n_2}},
\end{equation}
and $G_s(t) = G_c(t-\tau_d)$. Here, $\tau_d$ is the delay between the center and surround responses. All other parameters are constants derived from model fits in \citet{cai1997spatiotemporal}.

A popular choice for modeling the spatial component of thalamic relay cell RFs is the difference-of-Gaussians (DOG) model. We model the spatial component of the LGN RF in this way when there is no cortical feedback. Center and surround spatial profiles are Gaussian functions; for example, the center spatial filter is given as:
\begin{equation}\label{eq:spatial}
F_c(x) = A_{c} e^{-x^2 / \sigma_{c}^2},
\end{equation}
where $A_{c}$ is the amplitude of the center response and $2\sigma_{c}$ is the size of the RF center. The surround spatial profile $F_{s}(x)$ is defined analogously. When cortical feedback is included, spatial profiles are modeled according to the extended difference-of-Gaussians (eDOG) model, described in more detail in the next section.

\begin{table*}
\centering
\begin{tabular}{| c | c  | c  | c || c | c | c || c |}
\hline
\textbf{Symbol} & \textbf{Physical meaning} & \textbf{Units} & \textbf{Eq. Ref.} & \textbf{Adult} & \textbf{8 week} & \textbf{4 week} & \textbf{Range}\\
\hline
$2n$ & Num. LGN cells per V1 cell & - &\ref{eq:input}& 40 & 40 & 40 & - \\
\hline
$W$ & Inhibition weight & - &\ref{eq:input}  & 1.25 & 1.25 & 1.25 & -\\
\hline
$\tau$ & Inhibition time delay & ms & \ref{eq:input} & 4 & 4 & 4 & - \\
\hline
$f$ & Stimulus frequency  & c/deg &\ref{eq:stim} & - & - & - & [0.001, 1.5]\\
\hline
$\phi$ & Stimulus phase  & rad & \ref{eq:stim} & - & - & - & $\{0,\frac{\pi}{2},\pi,\frac{3\pi}{2}\}$\\
\hline
$K_1$ & - & - & \ref{eq:temporal} & 1.05 & 1.05 & 1.05 & -\\
 \hline
$K_2$  & - & - &  \ref{eq:temporal} & 0.7 & 0.7 & 0.7 & -\\
 \hline
$c_1$ & - & - &\ref{eq:temporal} & 0.15 & 0.15 & 0.15 & -\\
 \hline
$c_2$  & - & - &\ref{eq:temporal}& 0.1 & 0.1 & 0.1 &-\\
 \hline
$t_1$  & - & ms &\ref{eq:temporal}& -6 & -6 & -6 &-\\
 \hline
$t_2$  & - & ms & \ref{eq:temporal}& -6 & -6 & -6 &-\\
 \hline
$n_1$  & - & - & \ref{eq:temporal}& 7 & 7 & 7 &-\\
 \hline
$n_2$  & - & - &\ref{eq:temporal}& 8 & 8 & 8 &-\\
 \hline
$\tau_d$ & Center-surround delay & ms &\ref{eq:temporal} & 8 & 12 & 16 & [5, 20]\\
\hline
$\sigma_c$ & Center size & deg & \ref{eq:spatial} & 0.4 & 0.4 & 0.4 & [0.2, 1]\\
\hline
$\sigma_s$ & Surround size & deg & \ref{eq:spatial} (related) & 1.0 & 1.0 & 1.0 & [0.5, 1.5]\\
\hline
$A_s / A_c$ & Relative surround strength & - & \ref{eq:spatial} (related)  & 0.3 & 0.2 & 0.1 & [0, 1]\\
\hline
\hline
$\STC$ & Space-time constant & ms/deg$^2$ & \ref{eq:stc}  & -3.5 & -3.5 & -3.5 & - \\
\hline
$d_m$ & Median center size & deg &\ref{eq:stc} & 1.15 & 1.15 & 1.15 & - \\
\hline
\hline
$C$ & Cortical feedback strength & - &\ref{eq:dog2} & 0 & 0 & 0 & [-0.75, 0.75]\\
\hline
$a$ & Cortical feedback spread & deg &\ref{eq:dog2}  & 0.075 & 0.075 & 0.075 & -\\
\hline 
\end{tabular}
\caption{Typical parameter values for each age group considered. Parameters with any biological relevance (i.e., those that are not just a consequence of model fits to data) are listed by name. If the parameter was varied in a numerical experiment, a range of values used is also given. More details on parameter choice can be  found in Section \ref{ss:param}.}
\label{tab:param}
\end{table*}

\subsection{Feedback connections}\label{ss:feedback}

We do not incorporate cortical feedback connections explicitly, but rather by modifying LGN spatial profiles according to the extended difference-of-Gaussians (eDOG) model. A nice feature of this model is that (given certain model choices, described in more detail below) it reduces to the familiar difference-of-Gaussians model (up to a scaling constant) when cortical feedback parameters are set to 0.

In this section, we provide a brief overview of the eDOG framework and motivate our specific choice of model structure. For further details, we refer the reader to \citet{einevollextended}. The eDOG model is a firing-rate based model that incorporates corticothalamic feedback implicitly via modifications to the LGN receptive field structure. While this results in a more mathematically involved formulation for the RF than the traditional DOG, our model still maintains its simplicity, namely a low number of parameters (many of which can be attributed with a physiological meaning). 

The eDOG model is a formulation for LGN RFs that directly derives from a mechanistic model for an impulse-response function (here, interchangeable with the term \emph{receptive field}; see, e.g., Equation \ref{eq:lgnrf}) for the thalamic relay cell. The model aims to account for feedback effects to the LGN from a set of orientation-selective cortical populations, under a certain set of assumptions. 

Firstly, all cortical populations are considered to receive direct input from the thalamus and, in turn, provide direct feedback. Though it is true that the dominant input to the cortex from the LGN is in layer IV, there is also evidence of direct geniculate input to layer VI, where corticothalamic feedback is known to originate \citep{sherman2001exploring}. Furthermore, experiments by \citet{grieve1995differential} have shown that simple cells make up the majority of layer VI projections to the LGN. 

Secondly, the sum of feedback contributions from all cortical populations considered is assumed to be circularly symmetric. This is to say that the feedback effects are independent of the absolute orientation of the stimulus gratings. While the feedback contribution of a single cortical population is certainly anisotropic due to the orientation-selectivity of cortical cells, the net feedback effect can be expected to be (close to) circularly symmetric, since cortical populations cover all orientation angles. This idea is in accordance with experimental observations of corticothalamic feedback \citep{cudeiro1996spatial}.

The final assumption made in the derivation that we specifically mention is that corticothalamic feedback is assumed to have a ``phase-reversed, push-pull'' organization. This is to say that, for example, inhibitory feedback from a cortical ON-cell onto an LGN cell is accompanied by excitatory feedback from a cortical OFF-cell \citep{wang2006functional}. 

Despite these assumptions, the eDOG model has been shown to reproduce results from several experimental studies quite accurately. For a more complete exposition of all assumptions made in the derivation of the eDOG model, as well as validation of the model against experimental results, we refer the reader to \citet{einevollextended}. 

Under this set of assumptions, a general expression for the impulse-response function of an LGN cell accounting for cortical feedback can be derived (see Equation 26 from \citet{einevollextended}, which is analogous to a Fourier transformed version of our Equation \ref{eq:lgnrf}). Then, an expression for the space-time receptive field (directly analogous to our Equation \ref{eq:lgnrf}) can be generated by taking the inverse Fourier transform. 

Within the eDOG framework, the spatial component $F^*_{c,s}(x)$ of the receptive field for an LGN cell accounting for the effects of cortical feedback is described as:
\begin{equation}\label{eq:dog1}
F^*_{c,s}(x) = \frac{1}{(2\pi)^2}\int{e^{ikx}\frac{A_c e^{-k^2\sigma_c^2/4} - A_s e^{-k^2\sigma_s^2/4}}{1-Ce^{-k^2 a^2/4}}} dk.
\end{equation}
Here, the following parameters are defined similarly as in previous sections: $A_c$ and $A_s$ are the amplitudes of center and surround responses, $2\sigma_c$ and $2\sigma_s$ are the sizes of these responses. Additionally, $k$ is the wavenumber corresponding to the spatial frequency $f$ by $k = 2\pi f$. The feedback parameters $C$ and $a$ determine the strength and spatial spread of the feedback, respectively. See Table \ref{tab:param} for typical values of these parameters.

We choose a circularly symmetric Gaussian function to represent the kernel corresponding to the corticothalamic feedback loop (seen in the denominator of Equation \ref{eq:dog1}; for other options, see \citet{einevollextended}). Because there are not many physiological results guiding this decision, we make this choice primarily for mathematical convenience. 

In particular, this modeling choice has several nice properties, due to the following approximation. Using a series expansion, $1/(1-y) = \sum_{m=0}^\infty{y^m}$, with $y = C e^{-k^2a^2/4}$ and replacing the denominator term of the above integral expression, we arrive at our equation of LGN spatial profiles:
\begin{equation}\label{eq:dog2}
F^*_{c,s}(x) = \sum_{m=0}^\infty{C^m\left(A_c \frac{e^{-x^2/(\sigma_c^2 + ma^2)}}{\pi(\sigma_c^2 + ma^2)} - A_s \frac{e^{-x^2/(\sigma_s^2 + ma^2)}}{\pi(\sigma_s^2 + ma^2)}\right)}.
\end{equation}

This equation is an infinite sum of DOGs, and the first term of this sum ($m=0$) corresponds to the traditional feedforward model. The following terms correspond to ``corrections'' of this direct term due to $m$ rounds of the corticothalamic loop. We truncate this sum at the twentieth-order term, since higher terms are essentially equal to 0 for all values of $\left|C\right|$ we consider. This series uniformly converges for $|C| < 1$; see, e.g., Appendix 1 in \citet{einevollextended}.

The formulation in Equation \ref{eq:dog2} is especially convenient for two reasons. (1) It is easily related back to the traditional DOG model described in Section \ref{ss:lgn}. In particular, this form of the eDOG model reduces to the DOG model (up to a constant scaling factor) when feedback parameters ($C$ and $a$) are set to 0, as well as when only the first term of the series (corresponding to 0 rounds of the corticothalamic loop) is considered. (2) The structure of Equation \ref{eq:dog2} allows for easy separation of center and surround components for construction of the LGN RF according to Equation \ref{eq:lgnrf}. Specifically, the center response of the spatial LGN RF accounting for cortical feedback (analogous to Equation \ref{eq:spatial} without feedback) is:
\begin{equation}
F_c^*(x) = \sum_{m=0}^\infty C^m \left(A_c \frac{e^{-x^2/(\sigma_c^2 + ma^2)}}{\pi(\sigma_c^2 + ma^2)}\right),
\end{equation} 
with the surround component defined analogously.

In the following section, we describe how we measure and quantify model responses, as well as provide information as to our choice of parameters and implementation for the model described above.

\section{Analysis}\label{analysis}

\subsection{Quantifying the coarse-to-fine dynamic}\label{ss:quant}

\begin{figure}
\begin{center}
\includegraphics[width=0.4\textwidth]{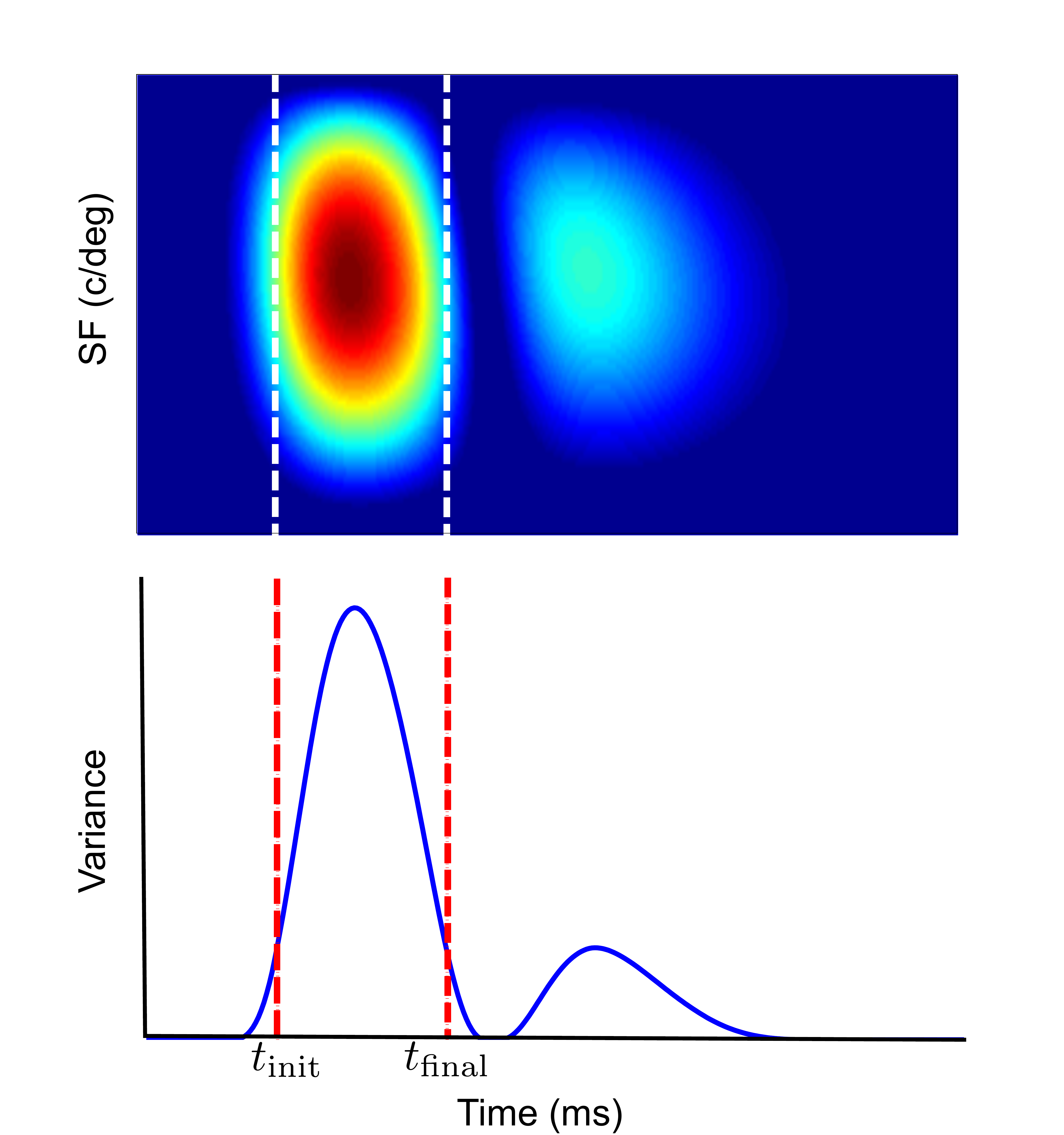}
\caption{Procedure to set the analysis window. Top panel is a representative spectrotemporal receptive field. Bottom panel shows the variance of the cell's response at every time slice. The boundaries of the analysis window $[t_{\text{init}}, t_{\text{final}}]$ are determined as the times where this variance rises and falls to 20$\%$ of the maximum value. This corresponds roughly to the first phase of the LGN response. Dotted lines (white for the top panel, red for the bottom panel) mark the boundaries of the analysis window for the cell shown.}
\label{fig:analwin}
\end{center}
\end{figure}

We are interested in measuring the effects of various model parameters on dynamic SF tuning.  The magnitude of a cell's response to various spatial frequencies forms a tuning curve which peaks at a certain frequency $\SF_{\text{peak}}$. For a set of stimuli $S(f,\phi_i,t)$ with phases $\{\phi_i\}_{i\in I}$:
\begin{equation}\label{eq:sfpeak}
\SF_{\text{peak}}(t) = \argmax\limits_{f} \overline{R}(f,t),
\end{equation}
where 
\begin{equation}\label{eq:avgR}
\overline{R}(f,t) = \frac{1}{|I|} \sum_{i\in I} R(f,\phi_i,t).
\end{equation}
Our choice of phases $\{\phi_i\}_{i\in I}$ is given in Table~\ref{tab:param}, and recall that $R(f,\phi,t)$ is defined in Equation \ref{eq:cortresp}. Dynamic SF tuning refers to the changes in time of a cell's maximal frequency during exposure to a visual stimulus. 

We choose to use the shift in peak SF as a metric for the strength of the spatial coarse-to-fine process. This is a fairly intuitive choice: cells displaying a strong coarse-to-fine response will naturally produce large shifts, as the cell preferentially responds to low frequencies at early timepoints and to high frequencies at later timepoints.

We formally define SF shift as the log ratio of $\SF_{\text{peak}}$ values measured at the final ($t_{\text{final}}$) and initial ($t_{\text{init}}$) time points of a predetermined analysis window:
\begin{equation}
\Delta \SF = \log_2 \frac{\SF_{\text{peak}}(t_{\text{final}})}{\SF_{\text{peak}}(t_{\text{init}})}.
\end{equation}

Here the boundaries of the analysis window, $t_{\text{final}}$ and $t_{\text{init}}$, are calculated as the times at which the variance of SF response rises and falls to 20$\%$ of the maximum value. This time period corresponds roughly to the first phase of the LGN response \citep{allen2006dynamic}. See Figure \ref{fig:analwin} for an example. 

Cortical peak spatial frequency is directly observed from the responses of simple cells to stimuli, averaged across all four phases. These computations are analogous to those used in previous studies on cortical SF tuning \citep{bredfeldt2002dynamics,frazor2004visual}. 

 
\subsection{Parameter choice}\label{ss:param}

We are interested in comparing the cortical coarse-to-fine dynamic at various stages of development. In order to do this, we measure simple cell responses to input from three LGN cell populations: one representative of kittens at 4 weeks postnatal, one of kittens at 8 weeks postnatal, and one of mature cats. All parameter values corresponding to these populations are either directly taken or estimated from previously published, publicly available data. In this section we detail and reason our choice of individual parameter values.

We choose spatiotemporal RF parameters based on model fits based on experimental data. Values for $t_1$, $t_2$, $W$, $\tau,$ and $\STC$ were taken from~\citet{allen2006dynamic}. Values for $K_1$, $K_2$, $c_1$, $c_2$, $n_1$, $n_2$, and $A_s/A_c$ are fixed as the geometric means of their distributions from \citet{cai1997spatiotemporal}. For parameters $\sigma_c$ and $\sigma_s$, we truncate the tail of the distribution before estimating the value.  Because the distributions for $\tau_d$ were not given explicitly, we chose values which we felt best qualitatively recreated the RFs shown in \citet{cai1997spatiotemporal}, Figure 3. Adult and kittens at 8 weeks postnatal $\tau_d$ were taken to be somewhat higher than, but well within one SD of the (arithmetic) mean given in the text; the value for kittens at four weeks postnatal was fixed at the average value given. If values for a parameter were not found to vary significantly between age groups, values for all developmental stages were fixed using adult data.

Because there have been few conclusive experimental results regarding corticothalamic feedback in the visual pathway, values for parameters corresponding to cortical feedback were chosen in a less exact manner.   Values for cortical feedback strength $C$ were chosen so as to lie within the constraints of the approximation described in Section~\ref{ss:feedback} (namely $|C| < 1$).The value for spatial spread, $a=0.075$~deg, was chosen to qualitatively support results involving corticothalamic feedback---namely, that feedback serves to strengthen the antagonistic surround response \citep{alitto2003corticothalamic,briggs2008emerging,andolina2012effects} but does not have a large effect on the size of the surround response \citep{andolina2012effects}.

A table of parameter values used for each age group, as well as a range of values for parameters which were varied, is presented in Table~\ref{tab:param}.




\subsection{Randomization of LGN population parameters}\label{ss:random}

In addition to characterizing the effects of certain parameters on spatial coarse-to-fine processing, we also wish to demonstrate the robustness of the model to variability in LGN population parameters. For this, we perform numeric experiments with randomized values for parameters. All properties of the LGN cell population are in accordance with measurements reported by \citet{alonso2001rules}.  More details on the motivation behind these values can be found in \citet{allen2006dynamic}.

Specifically, the position of an LGN cell is drawn from a normal distribution with mean equal to the center of the cortical subregion and standard deviation 0.15$^\circ$. RF center sizes $d_c$ are drawn from a normal distribution with mean 0.8$^\circ$ and SD 0.6$^\circ$, rectified $< 0.7^\circ$. In our randomized population, we simply shift the temporal profile along the time axis by shifting the parameters $t_1$ and $t_2$ (Equation \ref{eq:temporal}, in Section~\ref{ss:lgn}) by a function of the size of the RF~\citep{weng2005receptive}:
\begin{equation}\label{eq:stc}
\text{shift} = \text{STC} \left[\pi\left(\frac{d_c}{2}\right)^2 - \pi\left(\frac{d_m}{2}\right)^2\right],
\end{equation}
where $d_c$ is the center diameter of the thalamic cell, and $d_m$ is the median center diameter \citep{allen2006dynamic}. We multiply these values by a space-time constant ($\STC = -3.5$ ms/deg$^2$) to obtain a population of LGN cells with latencies similar to experimental values reported in \citet{alonso2001rules}.

\subsection{Implementation}

All equations were discretized and solved numerically using difference methods. For our randomized population, average values of SF shifts were calculated from 100 trials for each parameter set. 
 
All calculations were performed in MATLAB (R2012a Student Version, The MathWorks, Inc.). MATLAB scripts will be made available on the web at the following URL: www.ocf.berkeley.edu/$\sim$jnirody.



\section{Results}\label{results}

In this section we present the results of our analysis. We study how developmental changes in the structure of LGN spatiotemporal RFs affect dynamic SF tuning, as well as how the contribution of corticothalamic feedback to the coarse-to-fine dynamic evolves throughout development.

We organize the presentation of our results as follows. We begin by focusing on the effects of two key model parameters: relative strength of the LGN RF surround $A_s/A_c$ and cortical feedback strength $C$ (see Table \ref{tab:param} for more information). Because the simplicity of our model allows thorough study of the relationship between model parameters and responses, we also give an overview of the effects of varying other LGN RF structural parameters on coarse-to-fine processing. Finally, we provide evidence of the robustness of our model to the randomization of parameters corresponding to LGN population organization.
\begin{figure}
\begin{center}
\includegraphics[width=0.4\textwidth]{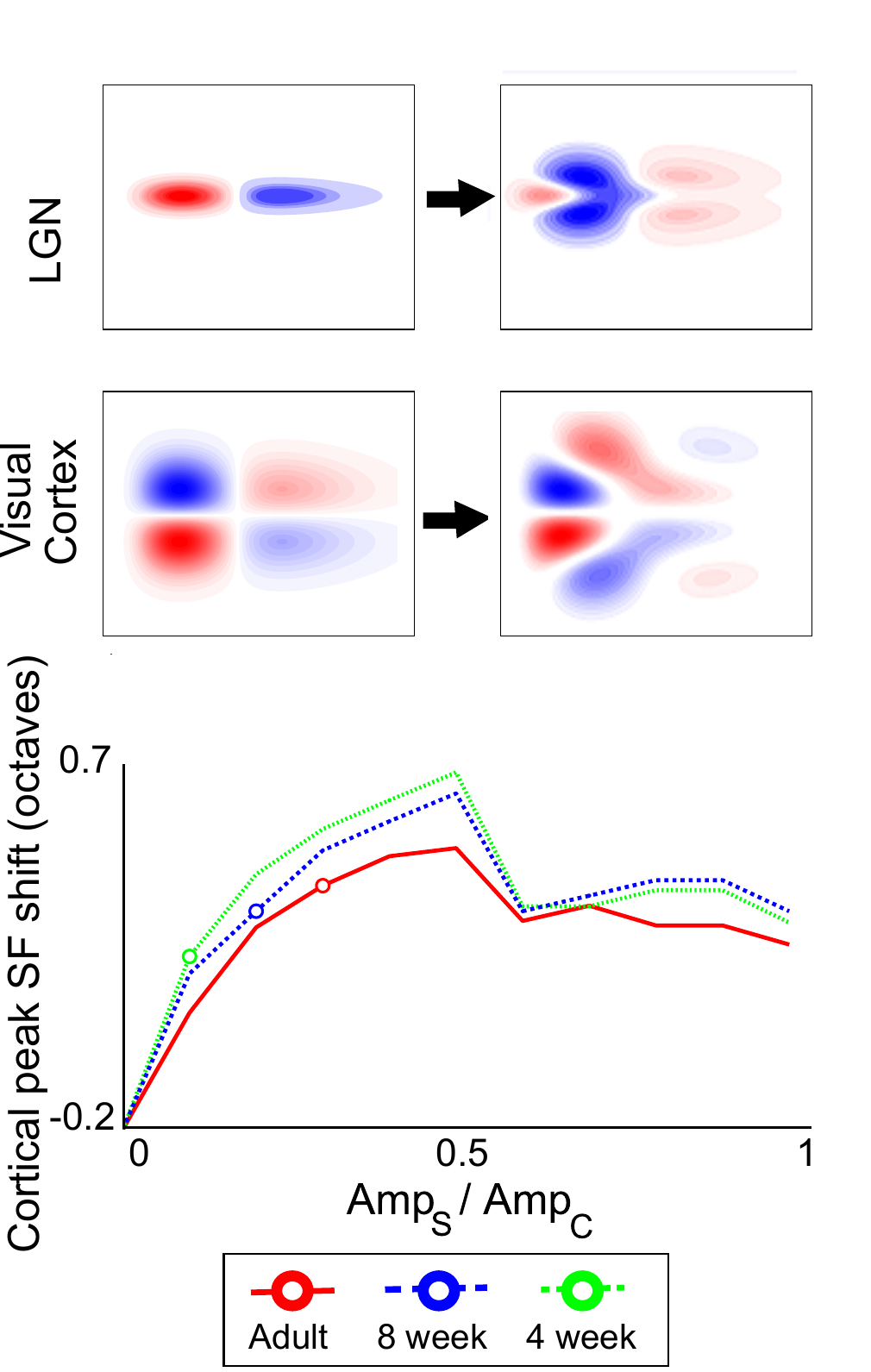}
\caption{Top panels show representative LGN and cortical spatiotemporal RFs at the low and high end of the interval for $A_s/A_c$ considered. Bottom panel displays cortical peak SF shift for various values of relative LGN surround strength $A_s/A_c$ for each age group. All other parameters are as in Table \ref{tab:param}.  Results for adults are shown in red, for 8-week old kittens in blue, and for 4-week old kittens in green. Circular markers are used to denote the responses for typical $A_s/A_c$ values used for each age group (which are listed in Table \ref{tab:param}).}
\label{fig:ssres}
\end{center}
\end{figure}

Out of the parameters involved in the structure of the LGN RF, we choose to concentrate our analysis on relative surround strength because it has been shown to vary significantly during development \citep{cai1997spatiotemporal}. We note that while the value of the center-surround delay $\tau_d$ has also been shown to change with age, the effects of varying this parameter were presented in \citet{allen2006dynamic}. Nevertheless, the effects of varying $\tau_d$ are briefly considered and reviewed in the context of development in Section \ref{ss:other}.

\subsection{Effect of relative surround strength}\label{ss:surrstr}

Both experimental and modeling studies have implicated the antagonistic relationship between center and surround responses in LGN cells in facilitating coarse-to-fine processing \citep{enroth1983spatio,allen2006dynamic}. In this section, we focus on model responses to changes in the relative strength of the LGN surround response, $A_s/A_c$. 

Figure~\ref{fig:ssres} shows the effects of various surround-center ratios ($A_s/A_c$) on cortical coarse-to-fine processing. In these calculations, all other parameters are set to the values listed in Table \ref{tab:param}. 

All three age groups exhibit a monotone increase in SF shift at low values of $A_s/A_c$. At the average values of $A_s/A_c$ reported by \citet{cai1997spatiotemporal} (as listed in Table \ref{tab:param}, $A_s/A_c = 0.3, 0.2, 0.1$ for adults, 8-, and 4-week old kittens, respectively), the shift in SF peak is 0.443  for adults, 0.396  for 8-week old kittens, and 0.314 for 4-week old kittens. These values are specifically marked on the plot. 

Interestingly, as the ratio $A_s/A_c$ increases, the peak SF shift for each age group reaches a maximum and begins to decline. Mature cats display a lower maximum SF shift (0.511) than both 8-week (0.610 ) and 4-week old kittens (0.649 ). This is likely due to the fact that kittens have a higher center-surround delay $\tau_d$ than adults (see Section \ref{ss:other}, Figure \ref{fig:other}A). 

This non-monotonic response is of particular note because it suggests that there is an optimal center-surround balance which maximizes SF shift. As the relative surround strength is generally lower in kittens (see circular markers in Figure \ref{fig:ssres}), mature cats on average display $A_s/A_c$ values which are relatively close to this ``optimum''. This suggests that strengthening of the antagonistic center-surround relationship (i.e., increasing the value of $A_s/A_c$) would have a proportionately larger effect on the peak tuning shift in kittens than in adults.

A mechanism which has been shown to heighten the antagonistic effect of the surround response is cortical feedback \citep{alitto2003corticothalamic,briggs2008emerging,andolina2012effects}. In the next section, we explore this hypothesis by comparing how the effects of cortical feedback on coarse-to-fine processing differ between age groups.

\begin{figure}[h!]
\begin{center}
\includegraphics[width=0.4\textwidth]{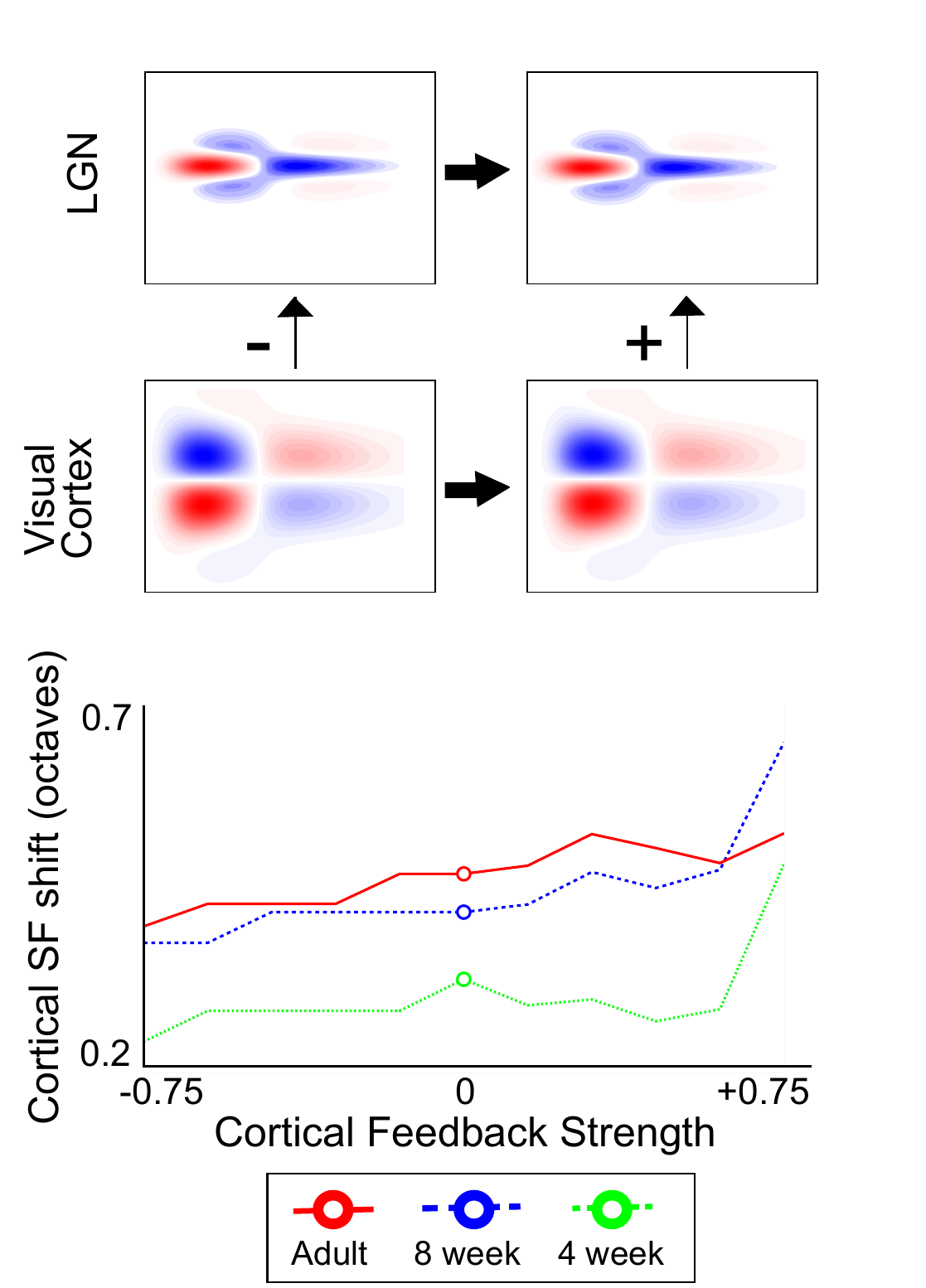}
\caption{As before, top panels show representative LGN and cortical spatiotemporal RFs. Bottom panel displays cortical peak SF shift for various values of cortical feedback strength $C$ for each age group. All other parameters are as in Table \ref{tab:param}.  Results for adults are shown in red, for 8-week old kittens in blue, and for 4-week old kittens in green. Circular markers are used to denote the responses when there is no cortical feedback (analogous to those marked in Figure \ref{fig:ssres}).}
\label{fig:corres}
\end{center}
\end{figure}

\subsection{Effect of cortical feedback}\label{ss:creed}

We also consider how cortical feedback affects spatial frequency tuning in these three groups. Though only excitatory corticothalamic connections exist, inhibitory feedback via thalamic interneurons has been observed \citep{alitto2003corticothalamic}. For this reason, we consider a variety of feedback strengths ranging from strongly negative ($C = -0.75$) to strongly positive ($C = +0.75$).

In Figure~\ref{fig:corres}, we study the effect of corticothalamic feedback on SF tuning shift. Unlike $A_s/A_c$, the model response to increasing values of $C$ is relatively monotonic: for all three age groups, as $C$ increases, so does the peak SF shift. However, we find that feedback is not equally effective in facilitating dynamic SF tuning in all age groups. 

Our results show that dominantly excitatory cortical feedback affects the coarse-to-fine process during development more strongly than in mature cats. When cortical feedback is mainly inhibitory ($C < 0$) or weakly excitatory ($C > 0$, $|C|$ small), there is a distinct difference in peak SF shift between age groups, with adults clearly showing the strongest response. However, as feedback becomes more strongly excitatory, this disparity is reduced. In fact, when $|C|$ is sufficiently high, the peak SF shift for 8-week old kittens surpasses that for mature cats. 

These results support the hypothesis we presented in the previous section: that a strengthening of the surround relative to the center (e.g., via cortical feedback connections) has a proportionally stronger effect on coarse-to-fine processing early in development.

\subsection{Effects of other thalamic RF parameters}\label{ss:other}

\begin{figure*}
\begin{center}
\includegraphics[width=\textwidth]{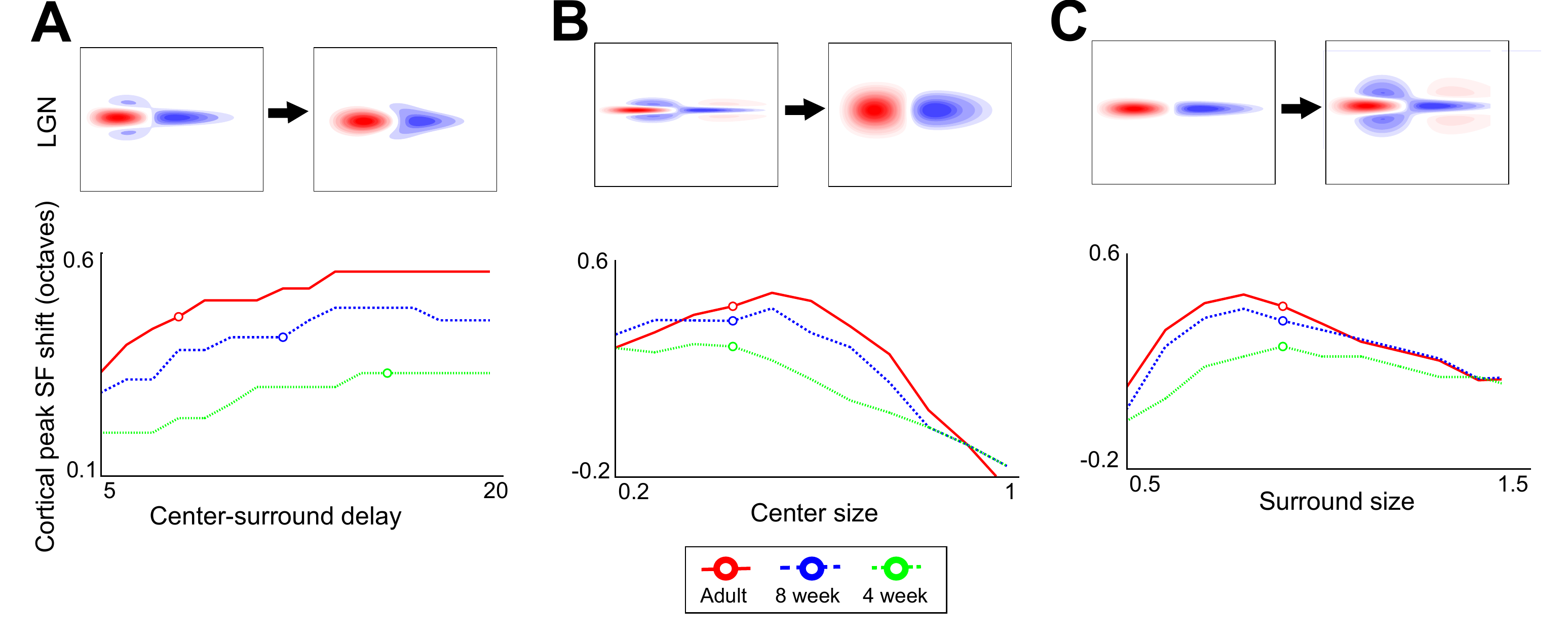}
\caption{Results for selected LGN RF structural parameters. Top panels show representative LGN RFs generated for parameters at either extreme. Bottom panels display cortical peak SF shift for each age group as parameter values are varied. All non-varied parameters are as in Table \ref{tab:param}. Results for adults are shown in red, for 8-week old kittens in blue, and for 4-week old kittens in green. As in other figures, circular markers are used to denote the responses for typical values of the parameters used for each age group (which are listed in Table \ref{tab:param}). \textbf{A,} Effect of center-surround delay $\tau_d$. \textbf{B,}~Effect of LGN RF center size $\sigma_c$. \textbf{C,} Effect of LGN RF surround size $\sigma_s$. Note that the limits of the y-axis in \textbf{A} are different from those in \textbf{B} and~\textbf{C}. }
\label{fig:other}
\end{center}
\end{figure*}

For the sake of completeness, we also present the relationship between the model response and other parameters related to the LGN receptive field structure. In Figure \ref{fig:other}, we show these results concerning (1) the delay between center and surround responses $\tau_d$, (2) the center size $\sigma_c$, and (3) the surround size $\sigma_s$. The results presented in Figure \ref{fig:other} emphasize the importance of the antagonistic relationship between center and surround responses in the LGN in driving cortical dynamic SF tuning. Note that with the exception of $\tau_d$, the values of these parameters do not significantly change during development \citep{cai1997spatiotemporal}. 

The relationship between center-surround delay $\tau_d$ and the coarse-to-fine process was explored in \citet{allen2006dynamic}, so we only provide a brief overview here in the context of development. Increasing the center-surround delay monotonically increases cortical peak SF shift (see Figure \ref{fig:other}A). This explains why peak SF shift is highest for 4-week-old kittens and lowest for adults at the same value for relative surround strength (see Figure \ref{fig:ssres}): the center-surround delay decreases during development, with kittens at 4 weeks displaying the longest delay and adults displaying the lowest (see Table \ref{tab:param}). 

In Figures \ref{fig:other}B and C, we show how RF center size $\sigma_c$ and surround size $\sigma_s$ affect coarse-to-fine processing. We explain these results in the context of their effects on the LGN center-surround relationship as follows. Both a moderately strong surround (but not too strong, see Figure \ref{fig:ssres}) and a high center-surround delay are necessary to facilitate dynamic SF tuning. Lowering or raising structural parameters out of an ``optimal range'' disrupts a balance which negatively affects the coarse-to-fine dynamic.

When the center size is too small, the surround seems to appear both too soon and too (relatively) strongly; see, e.g., the top left panel of Figure \ref{fig:other}B. However, the converse is far worse: the coarse-to-fine dynamic essentially disappears when the center is too large. The reason behind this is clear from observing the LGN RF: when the center diameter $2\sigma_c$ is significantly larger than the surround diameter (in Figure \ref{fig:other}B, surround diameter $2\sigma_s = 0.9^{\circ}$), it overpowers and stifles the surround response. Indeed, the model response drops off quickly when $\sigma_c > 0.45^{\circ}$ (see bottom panel of Figure \ref{fig:other}B). 

By the same reasoning, we see that a surround size significantly lower than $2\sigma_c$ (in Figure \ref{fig:other}C, $\sigma_c = 0.45^{\circ}$) also results in a similar-looking RF (top left panel, Figure \ref{fig:other}C). An overly large surround size also lowers the cortical peak SF shift. This is likely because an increase in the size of the response creates a ``diffusive'' effect, effectively lowering the center-surround delay (see top right panel, Figure \ref{fig:other}C). 

\subsection{Robustness to randomization of population parameters}\label{ss:robust}

\begin{figure*}
\begin{center}
\includegraphics[width=\textwidth]{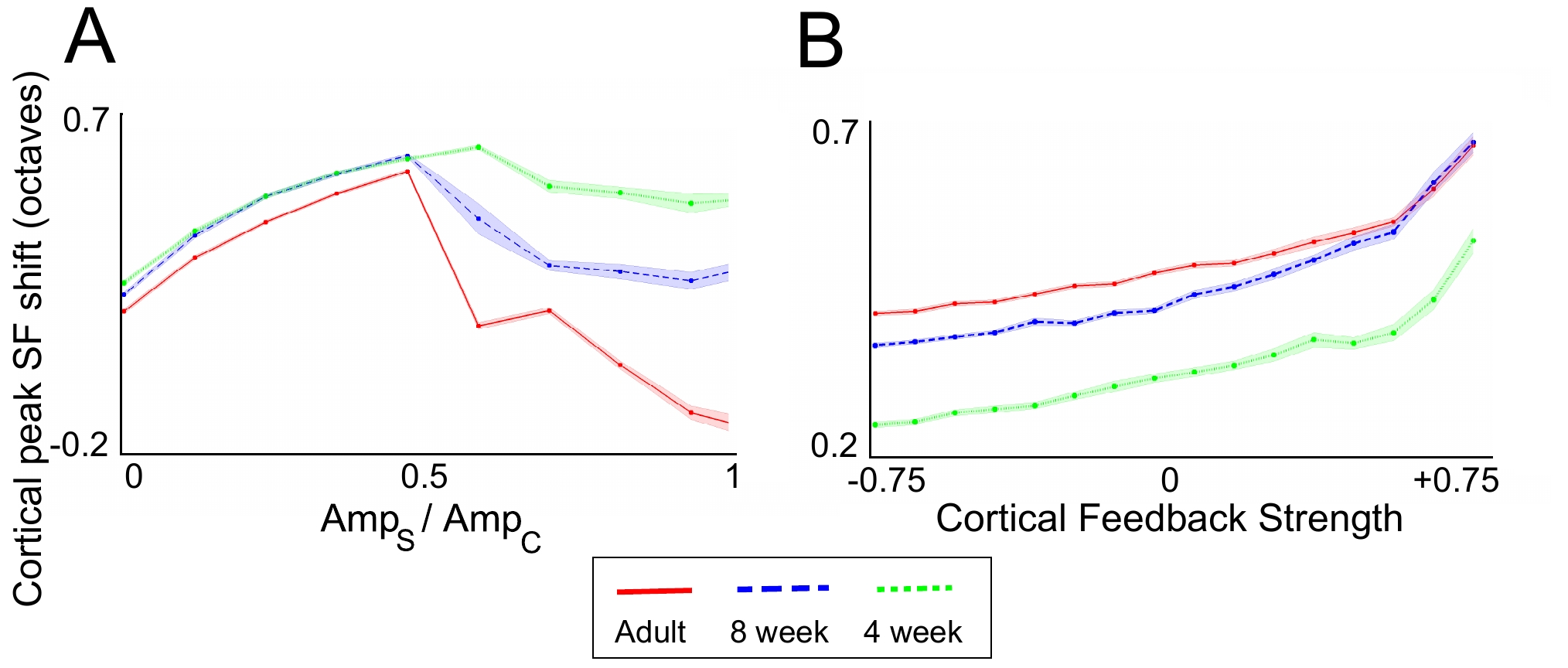}
\caption{Results for selected parameters when LGN population parameters are randomized. Each point corresponds to the mean taken from 100 simulations of our thalamocortical model, and shaded regions enclose $\pm$1 SEM away from the mean. Results for adults are shown in red, for 8-week old kittens in blue, and for 4-week old kittens in green. Parameters not being varied are as in Table \ref{tab:param}. \textbf{A,} Cortical peak SF shift as a function of surround-center ratio. Cortical feedback parameters are set to zero in this set of simulations. \textbf{B,} Cortical peak SF shift as a function of cortical feedback strength $C$. Note that the limits of the y-axis in \textbf{A} are different from those in \textbf{B}.}
\label{fig:robust}
\end{center}
\end{figure*}

Cortical cells receive input from a nonhomogeneous LGN cell population. For this reason, in addition to our parameter study, we demonstrate the robustness of our model to randomization of LGN population parameters. In Figure~\ref{fig:robust}A, we show the results of varying the relative surround strength (see Figure~\ref{fig:ssres} for results when population parameters are fixed). The qualitative results are unchanged: all three age groups exhibit a non-monotonic relationship between the parameter $A_s/A_c$ and model response, with the youngest age group showing the highest SF shift and the mature age group showing the lowest.

At the average values of $A_s/A_c$ reported by \citet{cai1997spatiotemporal} (see Table \ref{tab:param}), the mean peak shift was 0.520 for adults, 0.424 for 8-week old kittens, and 0.311 for 4-week old kittens. All values were significantly different from each other ($p < 0.001$ between 4-week old kittens and older age groups, $p < 0.05$ between 8-week old kittens and adults). 

Also as in Section \ref{ss:surrstr}, the peak SF shift for each age group in the randomized population reaches a maximum and begins to decline. Mature cats display a lower maximum SF shift (0.574) than both 8-week (0.600) and 4-week old kittens (0.616). 

In Figure \ref{fig:robust}B, we show results for the randomized LGN population when cortical feedback strength $C$ is varied. The qualitative results are unchanged from Section \ref{ss:creed}. All three age groups show increasing peak SF shifts as cortical feedback strength goes from strongly inhibitory to strongly excitatory. 

When $C$ is large and positive, the peak SF shift for 8-week old kittens is larger (though here, not significantly different) than for adults. Furthermore, changes in SF shift with respect to $C$ between age groups were found to be significantly different through a repeated-measures ANOVA at $\alpha=0.05$. These results confirm our observations when LGN population parameters are fixed. 

Results were also similar to those from non-randomized populations for other parameters considered (figures not shown).  

\section{Discussion}\label{discussion}

\subsection{Developmental changes in LGN RF structure}

We have presented a model to study how aspects of the LGN spatiotemporal receptive field affect the cortical coarse-to-fine dynamic, as well as how documented changes in the structure of these RFs impact this dynamic throughout the developmental process. Our results emphasize the importance of the antagonistic relationship between LGN center and surround responses in facilitating coarse-to-fine processing. Furthermore, the simplicity of our model allows us to investigate thoroughly the importance of specific features of the center-surround relationship.

We verify the results in \citet{allen2006dynamic} stating that higher center-surround delays result in a stronger cortical peak SF shift. This result is particularly important in the context of development because young kittens show significantly higher delays than adults \citep{cai1997spatiotemporal}. 

We also characterize the effects of other structural features of the LGN RF on cortical coarse-to-fine processing, which have not been explicitly considered previously. In particular, aspects of the spatial component of the RF, such as center and surround size and relative surround strength, play a large role in cortical SF tuning. Our results show that there is an ``optimal range'' for each of these parameters, suggesting that a delicate balance between the center and surround must be maintained for coarse-to-fine processing. Of special interest in studying the development of this process is the relative surround strength, which has been shown to increase during maturation.

Kittens in early developmental stages have relatively weak surrounds compared to adult cats. However, we find that, due to the high center-surround delays seen early in development, cortical peak SF shifts or kittens at 4 and 8 weeks postnatal are higher than would be expected (see circular markers, Figure \ref{fig:ssres}).

Another striking feature of our results concerning the effects of spatial RF parameters is the non-monotonicity of the model response. Here, we focus our discussion again on relative surround strength, as this parameter is the only one that has been shown to vary during development. Because kittens early in development show lower values of $A_s/A_c$ than adults, they are far from their ``optimal range'' for relative surround strength. This result indicates that a mechanism which enhances the center-surround antagonism would have a proportionally larger effect on young animals than on adults. 

This implication is particularly interesting because cortical feedback has been shown to strengthen the LGN center-surround relationship \citep{alitto2003corticothalamic,briggs2008emerging,andolina2012effects}. In the following section, we review our results regarding how cortical feedback affects cortical spatial coarse-to-fine processing in different stages of development.

\subsection{Role of cortical feedback during development}

Our results regarding the effect of LGN RF structural parameters suggest that cortical feedback may preferentially facilitate coarse-to-fine processing in kittens vs. in adults. Direct study via varying our parameter corresponding to cortical feedback verifies this: when cortical feedback is strong and dominantly positive, the peak cortical SF shift is higher for 8 week old kittens than for adults (see Figure \ref{fig:corres}).

Because of the clear importance of the recurrent connections to the thalamus from the cortex, there have been several experimental and computational studies involving the analysis of corticothalamic loops. However, a conclusion about the functional role of the corticothalamic pathway has yet to be reached. 

Experimental studies must precisely and specifically inactivate or remove corticothalamic connections in order to be able to make definitive claims about the nature of cortical feedback. Techniques such as lesions or ablation \citep{murphy1987corticofugal}, pharmacological blockage \citep{rivadulla2003receptive}, and transcranial magnetic stimulation \citep{de2007changes} have been used for this purpose. The results of these studies are neither in full agreement nor unequivocal in interpretation, possibly because these techniques tend to act on large areas of the cortex and cannot comment on more localized effects. 

Previous theoretical models have made use of the network structure of the corticothalamic circuit \citep{kohn1996corticofugal,worgotter1998influence, hayot2001modelling, yousif2007role}, making their results difficult to interpret in a physiologically relevant way due to the sheer number of parameters.

The advantage of a simple model is clearest here: the eDOG framework contains two biologically meaningful parameters corresponding to cortical feedback (feedback strength $C$ and spread $a$; see Table \ref{tab:param}). This very tractable number of parameters allows us to characterize how and why the effects of cortical feedback on coarse-to-fine processing change during development.  

\subsection{Model assumptions and limitations}

In this section, we review some assumptions and limitations of the model that may affect our specific analysis. The thalmocortical model we present in this paper is an integration of several previously published individual components. For this reason, we refer the reader to the papers for more details on more general limitations of these components and justification behind their assumptions. Specifically, the structure of the LGN and cortical populations are as given in \citet{allen2006dynamic}, and the eDOG model was presented initially by \citet{einevollextended}. A partial list of assumptions made in the eDOG model is also given in Section \ref{ss:feedback}.

As the direction of the excitatory-inhibitory balance of corticothalamic connections has not been fully characterized and is very unlikely to be static, we cannot conclude that a fixed set of parameters for this model will be universal. The inhibitory eDOG model has been shown to be in accordance with the experimental results of \citet{cudeiro1996spatial} and \citet{sillito2002corticothalamic}. These experiments, however, generated receptive fields using a different stimulus function (specifically, circular patch-gratings of various sizes). 

Other experiments involving inactivation of recurrent connections have found results in line with dominant excitation \citep{de2007changes}. Additionally, the parameters chosen for the excitatory eDOG model in this study qualitatively reproduce structural properties observed in RFs with an intact feedback circuit, namely an increased center size and a higher surround-center ratio when compared to RFs of decorticate animals \citep{andolina2012effects}. 

We also note in particular that the feedback term (modeled implicitly via modifications to the LGN RF structure, see Section \ref{ss:feedback}) does not change over time. Because the peak SF shift \emph{is} dependent on time, a model which incorporates dynamic cortical feedback would further elucidate the effects of these connections on coarse-to-fine processing. However, as there are very little conclusive experimental results on the nature of this feedback, our results are a useful first approximation. It is our hope that our results implicating cortical feedback in the development of coarse-to-fine processing will help direct deeper study (both computational and experimental) into this topic. We discuss this more explicitly in Section \ref{open}.

Finally, our firing-rate based model is largely conceptual, and does not include several aspects which would be implemented in a computational model, such as intracortical connections between simple cells with similar RF structure, non-deterministic firing rates, or spike thresholds. However, comparison of a similar conceptual model to more realistic simulations using integrate-and-fire neurons~\citep{troyer1998contrast} showed that while the addition of these mechanisms may affect exact numerical calculations, they do not alter not the qualitative results. We have omitted these mechanisms to reduce the number of free parameters and make the results easier to interpret.

\section{Conclusions and an open problem}\label{open}

We have presented a simple thalamocortical model which accounts for both feedforward and feedback connections in order to analyze the development of cortical coarse-to-fine processing. Our results characterize how specific aspects of the LGN RF structure affect cortical SF tuning, as well as implicate cortical feedback in the development of this process.

Additionally, our results put forward the hypothesis that the cortical coarse-to-fine process is strong very early in development (in the literature, 4 weeks postnatal is often the earliest time point measured for cats). Experiments measuring the cortical peak SF shift in different age groups would be ideal to test this hypothesis. Our results regarding cortical feedback could also be tested directly by comparing the coarse-to-fine dynamic of decorticate animals in several stages of development to control groups. We note however that care must be taken to use large sample sizes as there is a large variance observed for all parameters in kitten populations \citep{cai1997spatiotemporal}.

Until now, experimental results regarding the role of cortical feedback have been largely inconclusive. Because our results suggesting a specific potential role for cortical feedback in the visual system, they can be used to direct further, deeper study into elucidating the function of these connections.

\section*{Acknowledgements}
I acknowledge Ralph Freeman for the suggestion of studying the developmental process and for helpful discussions. I also thank Bartlett Moore for many helpful discussions and for reading over earlier versions of this paper. I am also very grateful to two anonymous reviewers whose thorough comments vastly improved this manuscript.

\bibliographystyle{spbasic}      
\bibliography{references}   

\end{document}